%% file: main.tex
\documentclass{article}

\usepackage{PRIMEarxiv}

\usepackage{cite}
\usepackage{amsmath,amssymb,amsfonts}
\usepackage{algorithm}
\usepackage{algpseudocode}
\usepackage{graphicx}
\usepackage{textcomp}
\usepackage{url}

\usepackage{hyperref}
\usepackage{bm}
\makeatletter

\title{BLS-MT-ZKP: A novel approach to selective disclosure of claims from digital credentials}

\author{
  Šeila Bećirović Ramić, Irfan Prazina, Damir Pozderac, Razija Turčinhodžić Mulahasanović, Saša Mrdović \\
  University of Sarajevo \\
  Faculty of Electrical Engineering \\
  Sarajevo\\
  Bosnia and Herzegovina\\
  \texttt{sbecirovic1, iprazina1, dpozderac1, rturcinhodzic, smrdovic@etf.unsa.ba}
}

\begin{document}
\maketitle

\begin{abstract}
Digital credentials represent a cornerstone of digital identity on the Internet. To achieve privacy, certain functionalities in credentials should be implemented. One is selective disclosure, which allows users to disclose only the claims or attributes they want. This paper presents a novel approach to selective disclosure that combines Merkle hash trees and Boneh-Lynn-Shacham (BLS) signatures. Combining these approaches, we achieve selective disclosure of claims in a single credential and  creation of a verifiable presentation containing selectively disclosed claims from multiple credentials signed by different parties. Besides selective disclosure, we enable issuing credentials signed by multiple issuers using this approach.
\end{abstract}


\input{Chapters/01Introduction}
\input{Chapters/02Preliminaries}
\input{Chapters/03RelatedWorks}
\input{Chapters/04CombinedApproach}

\input{Chapters/05Results}

\input{Chapters/06Discussion}
\input{Chapters/07Conclusion}

\bibliographystyle{unsrt}  
\bibliography{Bibliography/main}

\newpage

\end{document}

%% file: Chapters/01Introduction.tex
\section{Introduction}
\label{section1}
Digital credentials represent attestations, i.e. evidence of an individual's qualifications, claims, or achievements \cite{ramic}. They are the digital equivalent of ``paper'' credentials that people carry to prove their identity or qualification, e.g., identity card, driver's licence, passport and diploma. This paper focuses on this definition of digital credentials. The definition of digital credential should be distinct from the term used in other fields of computer science, where it can represent a simple password \cite{sedlmeir2021digital}. 

Since the term ``digital credential'' was introduced, an ongoing effort has been made to standardize it. David Chaum's theoretical construction of anonymous credentials from 1985 \cite{chaum1985security} represents the first significant development of digital credentials. These types of credentials enable users to prove that they possess credentials and can disclose information from them while maintaining anonymity  \cite{baldimtsi2013anonymous}. 
Further advancements followed, including the first practical implementation by Camenish-Lysanskaya \cite{camenisch2001efficient}, to the development of attribute-based credentials in IBM's Idemix \cite{camenisch2002design,uliknig_2023_identity} and Microsoft's U-Prove \cite{paquin2011u}. 
More recently, the focus shifted toward standardizing and developing verifiable credentials, which are tamper-evident and cryptographically verifiable. 
As an open standard, they are being increasingly developed to incorporate properties that enable preservation of privacy, such as selective disclosure and data minimization \cite{fett_2022_selective,_2019_hyperledger}. Users share information they must or want with other parties using selective disclosure. Verifiers require information from users for authentication and to provide them with services. With selective disclosure, users can  disclose only what is necessary and other information if they choose to. Selective disclosure represents a privacy-enhancing mechanism that has been extensively studied in recent years \cite{ramic}. 
Currently, the selective disclosure research area is expanding, with different approaches introduced regularly. There is no clear winner or the ``best'' universal solution for selective disclosure in digital credentials. It is therefore necessary to find a solution that satisfies all the requirements for selective disclosure schemes and legal regulations if needed.

This paper introduces a new approach to selective disclosure of claims in digital credentials. This approach represents a combination of different elements with a concrete implementation using the hash-based method of Merkle trees, the signature-based method using Boneh-Lynn-Shacham signatures and the zero-knowledge proof (ZKP) method Bulletproofs. The aim of this work is to create an approach to selective disclosure that fulfills the requirements defined by ETSI (The European Telecommunications Standards Institute) \cite{etsi}. With this approach, it is possible to selectively disclose claims from multiple digital credentials, combining them into one presentation with the ability to prove that claims belong to the user. This approach also enables multiple issuers to issue one credential and to have a user sign their credential, which complies with specific real-world examples. In addition, it enables proving that a claim value belongs to a specified range without revealing it. This is the first solution with all of these features. As far as we know there is no other approach that can achieve all selective disclosure requirements as defined in \cite{etsi}.

%% file: Chapters/02Preliminaries.tex
\section{Preliminaries}\label{sec2}
\subsection{Digital credential system and roles}
\begin{figure}[h]
    \begin{center}
        \includegraphics[width=1\columnwidth]{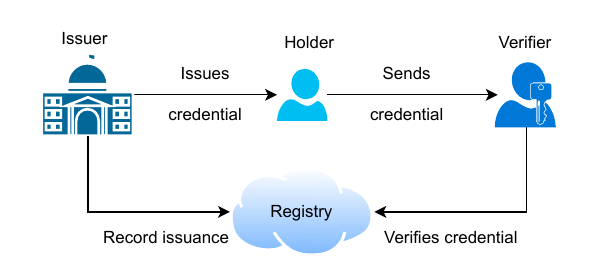} 
        \caption{Process of digital credential issuance and user roles}
        \label{fig:proces}
    \end{center}
\end{figure}

Digital credentials are electronic certificates or documents issued by an entity to verify an individual's qualifications, claims, or accomplishments. In a digital credentials system, users rely on these certificates because they are easy to manage and verify. 
The following explanation of credential system is based on a verifiable credentials system because it is the current standard.
The credential system has three roles: Issuer, Holder, and Verifier. Figure \ref{fig:proces} shows the process of issuance and verification alongside the user roles. Issuer issues and signs the credentials provided to the holder. The issuer saves the issuance record in a publicly available registry. Holder keeps their credentials and sends them as a presentation to verifier when needed.  Verifier checks the validity of the issued credential against the information in the publicly available registry \cite{36}.

In the context of non-selective verification, the validity of the credential is checked based on the digital signature of the entire credential. An open issue with credential verification is how to trust the credential when some claims or parts necessary for signature verification are missing during the selective disclosure. 

\subsection{Selective disclosure}

Selective disclosure is a mechanism that allows holders to reveal only information that they must and to whom they must. It also allows them to reveal only the information they want and to whom they want.
Selective disclosure is a mechanism designed to preserve the privacy of individuals and organizations. It serves as a privacy design pattern that enhances users' trust, privacy and security by enabling data minimization in terms of collected data, compliance with data regulation (General Data Protection Regulation (GDPR) \cite{voigt2017eu} and the California Consumer Privacy Act (CCPA) \cite{CCPA}) and determining who can access user data and under what conditions \cite{pattern,dock}.

The development of digital identity has led to the establishment of regulatory and legal frameworks. In 2014, the European Union Council introduced a regulation, eIDAS (Electronic Identification, Authentication and Trust Services), which supports secure cross-border transactions by establishing a digital identity and authentication framework. In 2023, a provisional agreement on updating and modifying eIDAS was reached called eIDAS2. This revised regulation outlines requirements for selective disclosure to ensure convenience and personal data protection, including minimizing its processing \cite{eu03}. 
Due to the necessity of selective disclosure, several standardization drafts are currently developed by ETSI \cite{etsi} and the European Blockchain Services Infrastructure (EBSI) \cite{ebsi}. ETSI defines the following requirements for selective disclosure \cite{etsi}:
\begin{enumerate}
\item The possibility that the user selectively discloses attributes so that these attributes appear to be part of an attestation/credential other than the one they were originally part of; 
\item The possibility to selectively disclose attributes from at least two separately issued attestations issued by the same issuer;
\item  The possibility to selectively disclose attributes from at least two separately issued attestations issued by different issuers; 
\item The possibility to selectively disclose attributes from a single attestation; 
\item The selectively disclosed attributes are unlinkable by means other than the information shared in the attribute over multiple sharing sessions of the disclosed attributes to at least two different verifiers who can collude and compare the attribute disclosures they have received;
\item  The selectively disclosed attributes are unlinkable by means other than the information shared in the attribute over multiple sharing sessions of the disclosed attributes to the same verifier; 
\item  The selectively disclosed attributes are unlinkable by means other than the information shared in the attribute over multiple sharing sessions of the disclosed attributes to at least one verifier who can collude with the credential issuer and show the attribute disclosures they have received;
\item  Whether or not the verifier, upon receipt of selectively disclosed attributes, can confirm that the attributes were issued to the same identity subject that is presenting the attributes (or to an authorized representative thereof) and to no one else; 
\item  Whether or not the verifier, upon receipt of selectively disclosed attributes, can confirm that the attributes describe the same identity subject that is presenting the attributes (or to an authorized representative thereof) and to no one else.
\end{enumerate}

These requirements define that the approach to selective disclosure must:
\begin{itemize}
\item support multi-show unlinkability (5, 6 and 7);
\item support combined presentations (2 and 3);
\item be resilient against colluding parties (5 and 7);
\item assure holder binding and proper pairing of presented disclosures (1, 4, 8 and 9).
\end{itemize}

Additional requirements can arise from observing the spectrum of privacy for digital credentials, which ranges from pseudonymous to strongly identified. Individuals have varying comfort levels regarding the information they are willing to share and the insights that can be inferred from it. This is especially true when dealing with sensitive data such as medical records \cite{ULLAH2023326}. A key selective disclosure element is understanding that sometimes only proving a value without revealing it is needed, e.g., proving that one's age is over 21 \cite{a2019_verifiable_data}.

From the definition of selective disclosure and the privacy spectrum, we explicitly define following additional requirements:
\begin{enumerate}
\setcounter{enumi}{9}
\item The disclosed attributes are part of a valid credential, and by using them, it is possible to validate the entire credential;
\item The disclosed attributes are part of a valid credential issued by an issuer whose origin (issuer and issuance) can be verified;
\item It is possible to prove the properties/values of attributes (range proofs, set membership) without disclosing them to protect privacy.
\end{enumerate}

The memory and time requirements for selective disclosure key components (issuing credentials, generating and verifying the presentation) are currently not defined. However, since selective disclosure must be performed on different agents, which means on different devices (computers, servers, mobile devices), it is necessary to find an approach to selective disclosure that can be performed on all of them. In addition, the approach's scalability is crucial for real-world use cases. Standardization of approaches must also consider potential changes in the post-quantum period \cite{etsi}.

We recognized three use case scenarios to evaluate the usability of the selective disclosure approach.

\begin{figure}[h]
    \begin{center}
        \includegraphics[width=\columnwidth]{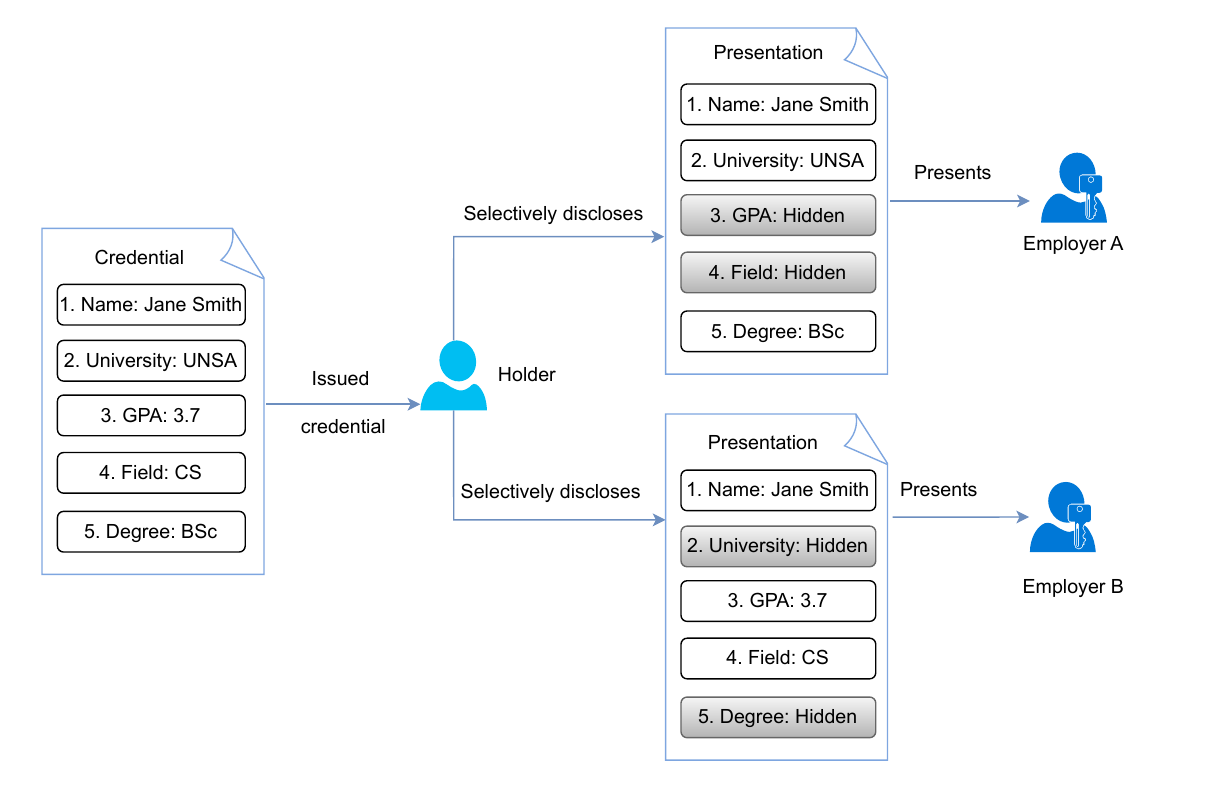} 
        \caption{Selective disclosure use case - single credential}
        \label{fig:scenario}
    \end{center}
\end{figure}

Figure \ref{fig:scenario} shows the main use case scenario of selective disclosure. For example, a university student receives their diploma as a digital credential. They send their diploma to potential employers when they apply for a job. Certain employers do not need all of the data in the diploma. Therefore, the student shares the university name and degree name with employer A while they share their grade and domain of study with employer B. 

\begin{figure}[h]
    \begin{center}
        \includegraphics[width=\columnwidth]{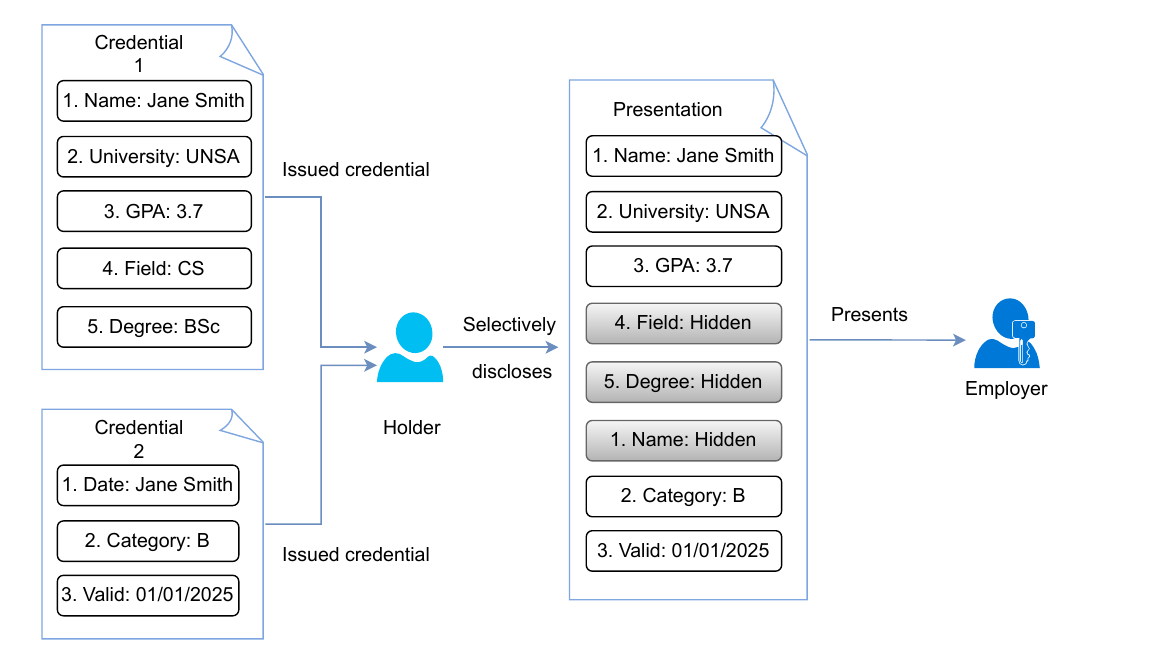} 
        \caption{Selective disclosure use case - multiple credentials}
        \label{fig:scenario2}
    \end{center}
\end{figure}

Another use case of selective disclosure is combining information from two credentials to send to an employer. The use case is shown in Figure \ref{fig:scenario2}. The user receives digital credentials for a driver's licence from the government and a diploma from the university. They combine these into one presentation and send it to employers. The claim about validity and category from the first credential and school and GPA from the other are added to one combined presentation.

\begin{figure}[h]
    \begin{center}
        \includegraphics[width=\columnwidth]{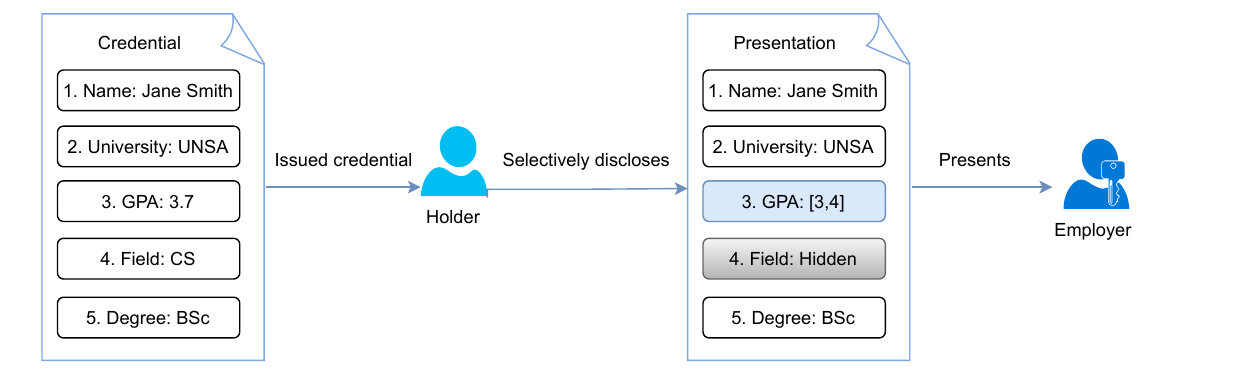} 
        \caption{Selective disclosure use case - zero-knowledge proof}
        \label{fig:scenario3}
    \end{center}
\end{figure}

The third use case requires proving something  has a value within required limits or in a required set without revealing that value. To achieve that, zero-knowledge proof is commonly used. For example, a student doesn't want to reveal their GPA, but they can derive evidence showing that the GPA is in the required range. The use case is shown in Figure \ref{fig:scenario3}. 

This use case is crucial in privacy-preserving contexts, i.e., when somebody needs to prove their age without revealing it, their salary range, or that they belong to a particular group.

In all use cases, employers should be able to confirm the validity of the presentations they receive using the public registry.

Different methods are used to achieve selective disclosure. Authors of \cite{a2019_verifiable} define the following methods:
\begin{itemize}
\item Atomic Credentials; 
\item Selective disclosure signatures;
\item Hashed values. 
\end{itemize}

Atomic credentials consist of a single claim, allowing users only to reveal the necessary claims/credentials during selective disclosure. Even though they are the most straightforward approach, atomic credentials are hard to manage. The number of credentials and overhead are increased. Additionally, there is no assurance that two claims can be correctly paired, which could result in presenting credentials with two unrelated or incorrect claims about the same subject. For example, two credentials about car type and car mileage from two different cars can be combined into one presentation. 

Atomic credentials option is generally discarded as a viable and applicable one for selective disclosure. They define an additional method, zero-knowledge proof, for selective disclosure, besides hash- and signature-based methods. They also show that a combination of methods can achieve selective disclosure. 

Hash-based methods hash claims about the subject in the credential. The user shares the presentation; they provide the hashed claims alongside the actual values of claims they wish to disclose. The verifier then hashes the provided values to verify that they match the original hashes. Merkle hash trees \cite{50,40,49}, hidden commitment schemes \cite{3,4,9}, and hashed message authentication codes \cite{22} are the most commonly used. Signature methods use particular types of signatures that allow the disclosure of specific claims. Most commonly used are CL-signatures \cite{camenisch2001efficient,camenisch2002dynamic, camenisch2003signature, camenisch2004signature,5} and BBS+ \cite{36} signatures. 
Zero-knowledge proof methods commonly implement zk-SNARKs (Zero-Knowledge Succinct Non-Interactive Argument of Knowledge) for achieving selective disclosure of claims \cite{24,32}.

Analyzing the existing approaches, we can see that the three methods and their combinations have certain advantages and disadvantages. Hash- and signature-based methods require showing the claims, while ZKP methods hide them/prove them. Both showing and hiding/proving the claims are necessary for an all-encompassing digital credential system. Hash methods are the least computationally intensive and least complex of the three, while ZKP methods are the most intensive and complex ones. Hash methods and ZKP methods alone cannot verify the issuer or holder, which is one of the key elements in digital credentials. Methods that use digital signatures usually add another layer of operations due to the canonicalization algorithm needed for claims. By combining specific methods, we can achieve what is lacking using only one method, but it can also add high computational costs and complexity. The approach should focus on fulfilling the requirements and enabling standard features, such as those in physical credentials, to achieve widespread adoption. 

The approach presented in this paper combines all three methods for selective disclosure to fulfill the requirements and scenarios. It combines Merkle trees, BLS signatures, and Bulletproofs to achieve selective disclosure. Below is a short description of each element to increase understanding. 

\subsection{BLS signatures}

The BLS (Boneh-Lynn-Shacham) signature scheme uses bilinear pairings for verification, with signatures represented as elements on an elliptic curve \cite{boneh2001short}. 
The scheme has the following characteristics \cite{boneh2003aggregate}:
\begin{itemize}
 \item Uniqueness and determinism: There is only one valid signature for any key or message:
\item Signature Aggregation: Multiple signatures created with different public keys for various messages can be combined into a single aggregated signature; 
 \item Threshold signatures: The scheme allows threshold issuing where multiple signers collaborate in producing a single signature.
\end{itemize}

The BLS signature scheme is provably secure in the random Oracle model and unforgeable under adaptive chosen message attacks. This security relies on the intractability of the computational Diffie-Hellman problem in a gap Diffie-Hellman group.

This scheme consists of three primary functions for generation, signing and verifying signatures \cite{boneh2001short}:
\begin{itemize}
\item generate - Algorithm for key generation selects a random integer $x$ such that $0<x<r$. The private key is $x$, and the corresponding public key is $g^x$, where $g$ is a generator of the group;
\item sign - Given the private key $x$ and a message $m$, the signature is computed by hashing the  message $m$, into a point on the elliptic curve $h=H(m)$ and then computing the signature as $\sigma=h^x$;
\item verify - To verify a signature, it is necessary to check whether the equation $e(\sigma,g)=e(H(m),g^x)$ holds where $e$ is the bilinear pairing function.
\end{itemize}

\subsection{Merkle tree}
\begin{figure}
    \begin{center}
        \includegraphics[width=\columnwidth]{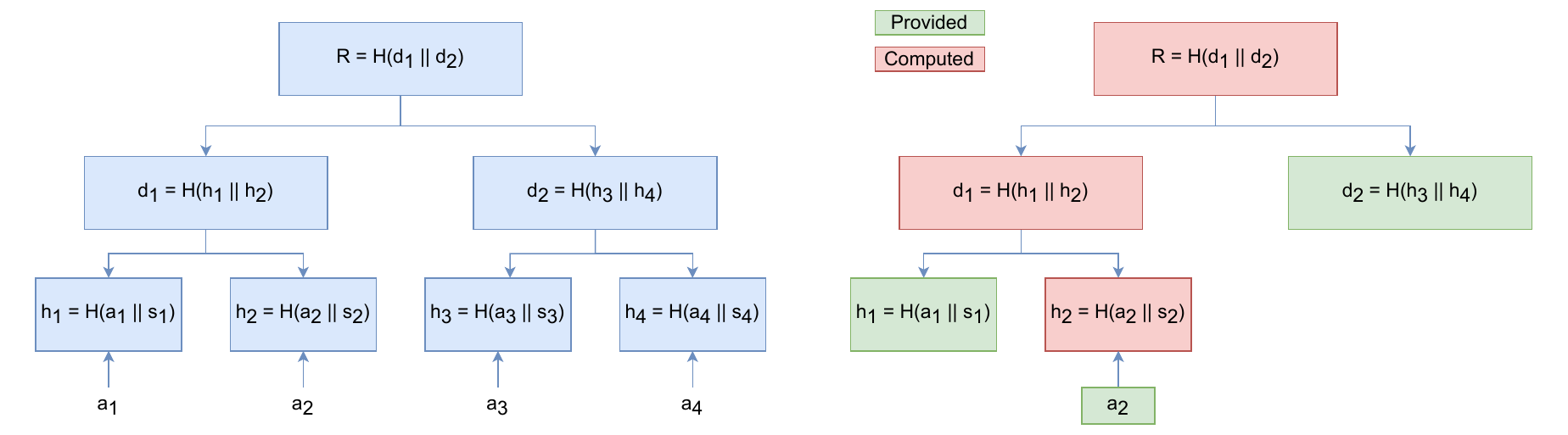} 
        \caption{Merkle tree generation and Merkle tree verification}
        \label{fig:merkle}
    \end{center}
\end{figure}

A Merkle tree, or hash tree, is a data structure that ensures data integrity and consistency for verification and synchronization. Named after Ralph Merkle, who patented it in 1979 \cite{merkle1982method}, the Merkle tree is a structure where each leaf node is labeled with the hash of data. Branch nodes are labeled with the hash of their child nodes. Merkle tree allows for efficient and secure verification of the contents of large datasets \cite{50}. Figure \ref{fig:merkle} shows the Merkle tree of elements and the verification method that an element belongs to the tree. Hashes for each attribute are calculated $h_i=H(a_i||s_i)$, where $s_i$ represents salt and is stored in the leaf nodes. The leaves root is determined by hashing pairs of nodes together $d_i = H(h_i || h_j)$. This process continues until the Merkle tree's root hash is obtained. To prove that element $a_2$ is part of the tree, the root is recreated using the element and its inclusion path  $[a_2,h_1,d_2]$. If the reconstructed root matches the original root, the verification is complete.

\subsection{Bulletproofs and Pedersen commitment}

A ZKP method is necessary for the third scenario, where Bulletproofs are utilized for range proofs.
Bulletproofs are a non-interactive zero-knowledge (NIZK) protocol designed for general arithmetic circuits, which produce compact proofs without a trusted setup. Bulletproofs are based on the discrete logarithm assumption. Using the Fiat-Shamir heuristic they are made non-interactive. The name comes from a description of its properties by one of the authors: ``Short like a bullet with bulletproof security assumptions'' \cite{bunz2018bulletproofs}. 

The core of the protocol is in its inner-product argument, initially presented by Groth \cite{groth2009linear} and later refined by Bootle et al. \cite{bootle2016efficient} 
This refinement provided proof for two unrelated binding vectors, Pedersen Commitments, that satisfy the inner-product relation. Building on this, Bulletproofs offer communication-efficient, zero-knowledge proofs, including range proofs derived from inner-product calculations on Pedersen commitments \cite{tarilab}.

A commitment scheme is a cryptographic primitive that allows a prover to commit to a chosen value without revealing it to the verifier (hiding) while ensuring the value cannot be altered (binding). It is widely used for creating blinded, non-interactive commitments \cite{pedersen1991non}. Bulletproofs are an efficient implementation that uses elliptic curve cryptography (ECC), called the Elliptic Curve Pedersen Commitment.
For a value  $x$, a random blinding factor $r$, a generator point $G$  and a point $H$ such that $H=x_HG$ ( where $x_H$ cannot be determined without solving the elliptic curve discrete logarithm problem), the commitment is calculated as $C(x,r)=xH+rG$. This commitment is homomorphic such that for messages $x, x_0$ and $x_1$, blinding factors $r$, $r_0$ and $r_1$ and scalar $k$, the following relation holds: $C(x_0,r_0)+C(x_1,r_1)=C(x_0+x_1,r_0+r_1)$ and $C(k\cdot x,k 
 \cdot r)=k \cdot C(x,r)$. This commitment is computationally binding and perfectly hiding \cite{tarilab}. 

In the Bulletproof protocol, the prover needs to convince the verifier that a Pedersen commitment $C(x,r)=xH+rG$ contains a value $x$ within the range $x\in[0,2^n-1]$. To achieve this, the prover represents $x$ as a vector of its binary bits $a=(a_1,…, a_n)$ where each $a_i$ is either 0 or 1. The core idea is to conceal all the bits in a single vector Pedersen Commitment. The proof involves showing that each bit $\omega$ satisfies $  \omega(\omega-1)=0$, confirming that $\omega$ is either 0 or 1 and that their sum equals $x$. Throughout the protocol, the verifier sends random linear combinations of constraints and challenges $ \in \mathbf{Z}_\mathbf{p} $ to the prover. In response, the prover constructs a vectorized inner product relation that contains vector $a$, the constraints and challenges $ \in \mathbf{Z}_\mathbf{p} $, and appropriate blinding vectors $ \in \mathbf{Z}_\mathbf{p}^n$. Because Pedersen commitments allow for a vector to be split and compressed,  the number of rounds for the inner product is reduced to a logarithmic number of rounds \cite{tarilab}.  
These elements are necessary for creating an approach to selective disclosure that fulfills the requirements. 

%% file: Chapters/03RelatedWorks.tex
\section{Related works}\label{sec3}

Merkle hash tree has been used to implement credentials with selective disclosure as demonstrated in several studies \cite{50, 40, 49}. In \cite{50}, authors use the Merkle tree for decentralized identity and store the root on the blockchain. Authors of \cite{40} use encryption on each attribute in order to prevent attack, while the authors of \cite{49} present a credential as a Merkle B+ tree where two nodes are for the same attribute (control node and data node). Besides the Merkle tree, other hashing techniques such as keyed-hash message authentication codes \cite{22,33} or hidden commitment schemes are used \cite{3,4,9}. These methods rely on hashing individual attributes. The attribute is revealed alongside salt during disclosure, while others are sent in their hashed form. On the other hand, Merkle tree do not require sending all of the attributes, but rather the proof. 

BLS signatures are a method used for selective disclosure, where each claim is individually signed, and an aggregated signature is created that represents the credential signature \cite{26}. Besides BLS signatures, the most commonly used signatures for selective disclosure are those based on special functions from elliptic curve cryptography \cite{6,21}, CL signatures \cite{5}, and BBS+ signatures \cite{36}. A key element of these signatures is that they require a canonicalization algorithm to be used for all attributes. This adds complexity to the solution and, in most cases, increases the size of the signature (depending on the number of attributes).

The ZKP method used in papers \cite{24, 32} is Groth-16, a zk-SNARKs method. This method requires a trusted setup, which adds additional complexity. It should be noted that other ZKP methods that do not require a trusted setup, such as Bulletproofs, are also used, while zk-STARKs methods are not commonly used in digital credentials \cite{oudesystematic}.

To meet specific requirements for selective disclosure, different researchers have combined various approaches. The following papers present combined approaches for selective disclosure. 
The authors of \cite{25} propose the CredChain architecture. This architecture integrates a hash-based method with redactable signatures. In this approach, each credential is represented by a Merkle tree with hashed attributes, and the root is signed with a redactable signature that includes both the signature and auxiliary information. During disclosure, the auxiliary data is updated in the signature. This method allows the credential to be signed once, generating multiple claims while minimizing user interaction and preventing correlation.

In \cite{39}, authors propose combining BLS signatures with hashing. The credentials attributes are hashed alongside the user's digital identifier (DID), which serves as a salt. Each claim is signed, and the aggregated signature is the signature of the entire credential.  

The Coconut scheme introduced in \cite{20} represents a selective disclosure credential scheme that combines threshold issuance signature scheme with ZKP for attribute proving.

Pointhcheval-Sanders Multi-Signatures (PS-MS) pairing scheme is used in paper \cite{31}. Using this approach, public key shares are aggregated into a single public key. This enables verification of a credential that includes ZKP of selectively disclosed values. 

The authors of \cite{10} expand the vector-commitments with zero-knowledge proof of knowledge (ZPKP) protocols to create a commitment, retrieve and update values. This method also involves blind signatures but has the drawback of being a one-time solution, meaning it requires re-issuance for each use. 

Structure-preserving signatures on equivalence classes (SPS-EQ) with fully adaptive NIZK proofs (AND and NAND proofs) is used in paper \cite{43} which allows signer-hiding as well. 

Merkle tree combined with Poseidon hash for zk-SNARKs is introduced in paper \cite{51}. This approach adds additional leaves to the tree, where the right half of tree contains attributes while the left represents metadata. Zk-SNARKs is used when needed. Otherwise, the Merkle tree is used in constrained devices. 

These methods usually combine two approaches: hash- and signature-based, ZKP and signature-based, or ZKP and hash-based to achieve additional functionalities. Each approach combines only two of the three. The combinations do not consider the requirements for selective disclosure defined in \ref{sec2} nor do they satisfy all of them (e.q. some approaches are focused only on unlinkability, while most of them do not consider combining credentials into one presentation). Combining all three approaches in a manner explained in the following section allows the fulfillment of all theoretical requirements and the implementation of practical solutions for selective disclosure without additional complexity.

%% file: Chapters/04CombinedApproach.tex
\section{Combined approach to selective disclosure} \label{sec4}
The primary purpose of the following approach is to present a solution for selective disclosure. 
Each cryptographic primitive used in this combined approach is chosen due to the specific requirement it can fulfill:
\begin{itemize}
\item A Merkle tree with different salts enables:
\begin{itemize}
\item Proving that disclosed attributes are part of a valid credential (requirements: 1, 4, 8, 9 and 10);
\item Validating the entire credential (requirements: 10);
\item Preventing the linkage of attributes with the same values to a certain degree (requirements: 5, 6 and 7);
\item Preventing recording of each attribute hash in a publicly available registry;
\item Creation of credentials without a fixed-size attribute lists for practical usage;
\end{itemize}
\item BLS signatures enable:
\begin{itemize}
\item Verification of the credential's origin, i.e., verification of the issuer (requirements: 11);
\item Verification of the identity owner(requirements: 1, 4, 8, 9);
\item  Verification of a presentation that consists of multiple credentials (requirements: 2 and 3);
\end{itemize}
\item Pedersen commitments with Bulletproofs enable:
\begin{itemize}
\item Proving attribute values (requirements: 12);
\item Achieving unlinkability through homomorphic values allows proving an attribute's existence in the tree without disclosing other information if required (requirements: 5, 6 and 7);
\item Smaller proof sizes that reduce communication costs;
\end{itemize}
\end{itemize}

All these methods were chosen as part of the approach due to their simplicity in implementation and the possibility of execution on computers or mobile devices.
 
The formalization of this approach is presented through the three main procedures: issuing credentials and generating and verifying the presentation. A formal definition of algorithms is given in \ref{alg:izd}, \ref{alg:pre} and \ref{alg:ve}. 

\begin{algorithm}[h]
\caption{Credential Issuance}\label{alg:izd}
\begin{algorithmic}[1]

\Require Set of attributes $\{a_i,v_i\}$, where $a_i$ is attribute name, $v_i$ attribute value, private key $sk$
\Ensure Issuance record $(R, \sigma)$, credential $(\{a_i\}, \{v_i\}, \{s_i\}, R, \sigma)$

\For{each attribute $a_i$ in the credential}
    \State Generate a randomized salt $s_i$
    \State Compute the Pedersen commitment $c_i = v_i \cdot G + s_i \cdot H$ on an elliptic curve, where $G$ and $H$ are base points
\EndFor
\State Construct a Merkle tree $\mathcal{M}$ with commitments $\{c_i\}$ as leaves
\State Compute the Merkle root $R$ of $\mathcal{M}$
\State Sign the root $R$ with the BLS signature using the private key $sk$ to obtain $\sigma = \text{BLS.Sign}(R)$

\State \textbf{Output:} 
\begin{itemize}
    \item \textbf{Issuance Record:} $(R, \sigma)$
    \item \textbf{Credential:} $(\{a_i\}, \{v_i\}, \{s_i\}, R, \sigma)$
\end{itemize}

\end{algorithmic}
\end{algorithm}
\begin{algorithm}[h]
\caption{Generating the Presentation}\label{alg:pre}
\begin{algorithmic}[1]

\Require Set of credentials $\{\text{cred}_j\}$, where each credential includes attribute names $\{a_i\}$, attribute values $\{v_i\}$, salts $\{s_i\}$, Merkle tree roots $R_j$, and signatures $\sigma_j$
\Ensure Presentation $(\{\text{proofs}_j\}, \sigma_{\text{agg}})$, where $\{\text{proofs}_j\}$ contains proofs and $\sigma_{\text{agg}}$ is the aggregated signature

\For{each credential $\text{cred}_j$ in $\{\text{cred}_j\}$}
    \For{each disclosed attribute $a_i$ in $\text{cred}_j$}
        \If{attribute $a_i$ requires a range proof}
            \State Generate range proof using Bulletproofs, denoted as $\text{proof}_{i-\text{range}}(c_i)$ on committed value
            \State Generate Merkle tree membership proof for the commitment $\text{proof}_{i-\text{Merkle}}(c_i, R_j)$
        \Else
            \State Generate Merkle tree membership proof $\text{proof}_{i-\text{Merkle}}(v_i, s_i, R_j)$, using attribute value $v_i$ and the salt $s_i$
        \EndIf
    \EndFor
\EndFor
\State Aggregate signatures from each credential to obtain $\sigma_{\text{agg}} = \text{Aggregate}(\{\sigma_j\})$

\State \textbf{Output:} 
\begin{itemize}
    \item \textbf{Presentation:} $(\{\text{proofs}_j\}, \sigma_{\text{agg}})$, where $\{\text{proofs}_j\}$ includes disclosed attributes with their salts or committed attributes and their corresponding proof: Bulletproof range proofs $\text{proof}_{i-\text{range}}$ and Merkle tree membership proofs $\text{proof}_{i-\text{Merkle}}$, and $\sigma_{\text{agg}}$ is the aggregated signature
\end{itemize}

\end{algorithmic}
\end{algorithm}
\begin{algorithm}[h]
\caption{Presentation Verification}\label{alg:ve}
\begin{algorithmic}[1]

\Require Presentation $(\{\text{proofs}_j\}, \sigma_{\text{agg}})$, where $\{\text{proofs}_j\}$ includes proofs (Bulletproofs, Merkle membership proofs) and $\sigma_{\text{agg}}$ is the aggregated signature, Issuance records $\{(R_j,\sigma_j)\}$, public keys $\{pk_j\}$
\Ensure \textbf{True} if verification succeeds, \textbf{False} otherwise

\For{each credential proof $\text{proof}_j$ in the presentation}
    \For{each disclosed attribute $a_i$ in $\text{proof}_j$}
        \If{attribute $a_i$ includes a range proof}
            \State Verify the range proof $\text{proof}_{i-\text{range}}(c_i)$ using Bulletproofs for commited attribute $c_i$
            \State Verify Merkle tree membership proof $\text{proof}_{i-\text{Merkle}}(c_i, R_j)$ for the commitment $c_i$
        \Else
            \State Verify Merkle tree membership proof $\text{proof}_{i-\text{Merkle}}(v_i, s_i, R_j)$ using attribute value $v_i$ and the salt $s_i$
        \EndIf
    \EndFor
\EndFor
\State Verify the aggregated signature $\sigma_{\text{agg}}$ for the presentation using public keys $\{pk_j\}$

\State \textbf{Output:} 
\begin{itemize}
    \item \textbf{True} if all Bulletproofs, Merkle tree membership proofs, and the aggregated signature are valid
    \item \textbf{False} if any verification step fails
\end{itemize}

\end{algorithmic}
\end{algorithm}

To understand these formal procedures, the building blocks of this approach are defined and evaluated through use cases.

\subsection{Creating a credential}

Digital credentials have predefined formats and should be treated as regular credentials with predefined fields and information available. They can be issued in JSON, XML, or other formats and sent in the same formats. The existence of a digital credential and verification method should be recorded in a publicly available registry, database, or DLT such as blockchain. 

\begin{figure}
    \begin{center}
        \includegraphics[width=\columnwidth]{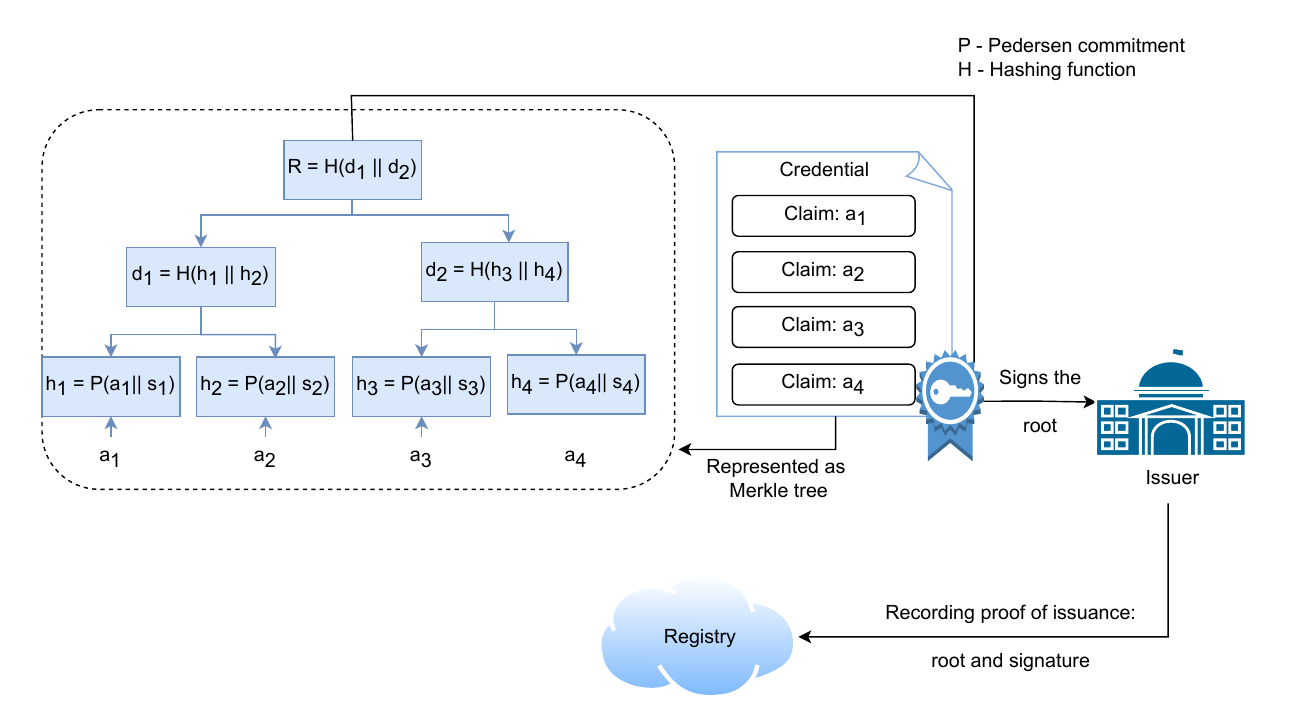}
    \caption{Issuance of credential}
    \label{fig:cred}
\end{center}
\end{figure}

In this approach, we propose presenting every digital credential as a Merkle hash tree. Depending on the format of the digital credential, each attribute has a corresponding leaf in the Merkle tree, as shown in Figure \ref{fig:cred}. 
The Merkle tree is generated using standard hashing functions. However, the first layer of leaves, created by hashing claims, uses a Pedersen commitment hash in this solution. The Pedersen commitment and the appropriate salt value allow for various ZKP circuits. One ZKP method used in this solution is Bulletproofs.
The root of the Merkle tree is signed using a BLS digital signature. The root and the signature are recorded in the previously mentioned public registry to have proof of issuance. Root hash represents a credential fingerprint, allowing a validity check of the credential while the signature authenticates the issuer. Due to the usage of signatures, public keys should also be publicly available and traceable to the issuer. Issuance is shown in Figure \ref{fig:cred}.

\begin{figure}
    \begin{center}
        \includegraphics[width=\columnwidth]{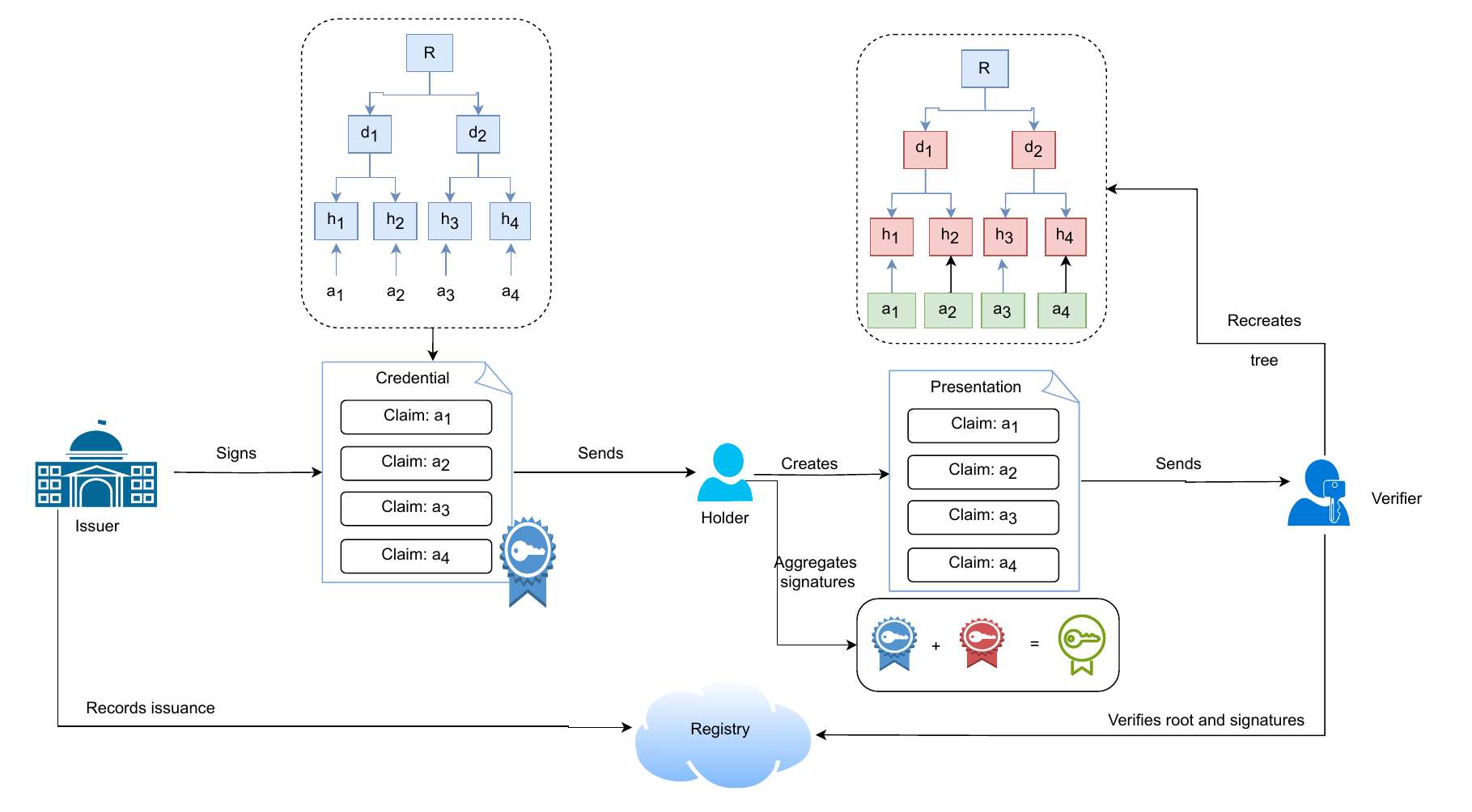}
    \caption{Regular use case of fully disclosing a single credential}
    \label{fig:reg}
\end{center}
\end{figure}

Holders can create verifiable presentations based on the credentials they have received. First, we will consider the regular use case of digital credential. This use case is shown in Figure \ref{fig:reg}. The holder creates a presentation based on a single credential. They sign the root of the credential using their private key. Afterward, they aggregate their signature with the issuer's signature into one single signature. This reduces the overall size of the signatures in the presentation and allows for verification of credential ownership. The aggregated signature is used to verify the issuer as the source of the credential and to whom it was issued. The root is used to verify the credential's validity. When the verifier receives the presentation, they recreate the Merkle tree to get the root. They check the root and the signatures against the public record of issuance and public keys.

\subsection{Multi-issuer credentials}
\begin{figure}
    \begin{center}
        \includegraphics[width=\columnwidth]{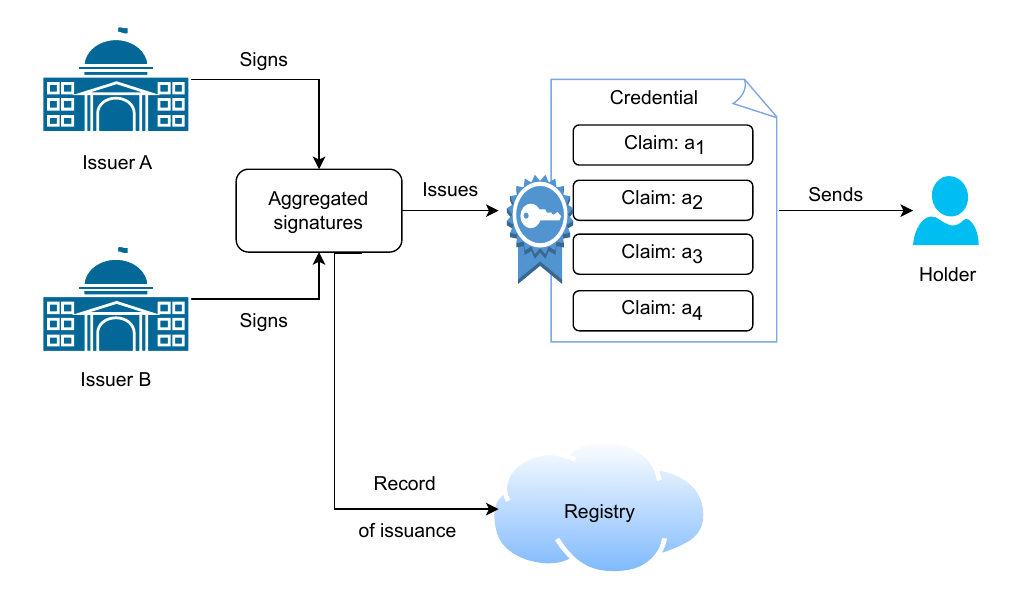}
 \caption{Issuing a certificate signed by multiple issuers }
 \label{fig:mulissuer}
\end{center}
\end{figure}

If credentials are created using the presented approach, creating different kinds of credentials is possible. One of the credentials that can be created is one signed by multiple issuers. One of the real-life examples where this is useful are university diplomas. Usually, a university diploma contains the signatures of the faculty's dean and the university's rector. Using the ability to aggregate signatures, several issuers can sign one credential while producing one signature. This aggregated signature is the same size as the signature of a single issuer. Aggregated signatures can be verified using all the required public keys. Figure \ref{fig:mulissuer} shows an example of issuing this type of credential.

\begin{figure}
    \begin{center}
        \includegraphics[width=\columnwidth]{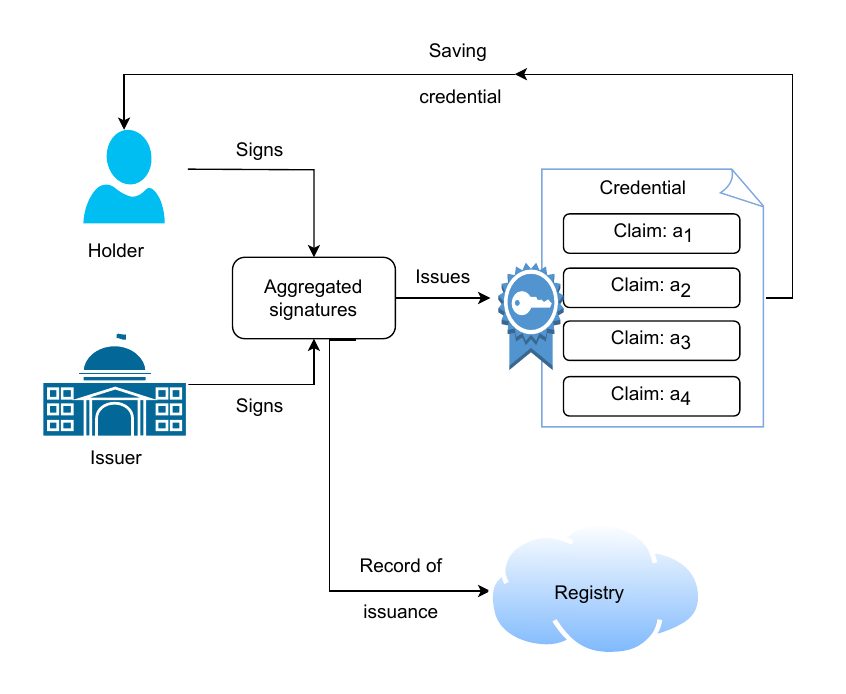}
 \caption{Issuing a certificate signed by the issuer and holder }
 \label{fig:usersign}
\end{center}
\end{figure}

In addition to the possibility of multiple issuers of one credential, it is also possible to have a use case where the holder can sign a credential alongside the issuer (shown in Figure \ref{fig:usersign}). There are specific situations where it is necessary to emphasize who the credential's owner is and if the appropriate holder received the credential, e.g., delegation cases. Therefore, it is possible to aggregate the signatures of the holder and issuer into one credential signature. To verify the credentials, signatures are checked using recorded public keys. 

\subsection{Selective disclosure of claims}

\subsubsection{Use case 1: Selective disclosure of claims from a single credential}

The first use case of selective disclosure of claims from a single credential is achieved primarily through Merkle trees. Figure \ref{fig:usecase1sol} shows this use case, and the exact steps are given below:
\begin{enumerate}
 \item Entities register their public keys in a public registry.
 \item In addition to the public key, the issuer registers the credential format to enable the correct generation of the Merkle tree.
 \item When issuing a credential, the issuer:
 \begin{enumerate}
 \item Creates a digital credential and signs the Merkle tree root using BLS.
 \item Records the issuance proof, i.e. its root and signature.
 \end{enumerate}
 \item The holder receives the credential and stores it in their digital wallet;
 \item The verifier requests presentation with specified claims;
 \item The holder sends their credential as a presentation to the verifier as follows:
 \begin{enumerate}
 \item They reveal claim values that verifier requested along with their salts; for the other values, they prepare the corresponding proofs generated using the Merkle tree. The presentation now consists of all the elements needed to recreate the root of the Merkle tree;
 \item They sign the root and aggregate their signature with the issuer's signature to create a unique presentation signature;
 \item They send the presentation to the verifier.
 \end{enumerate}
 \item When receiving a presentation, the verifier does the following:
 \begin{enumerate}
 \item Recreates the Merkle tree using the registered format, disclosed values, salts and the resulting hashes, thus obtaining the tree's root;
 \item Verifies the root using the public registry (thus validating the credential issuance). They verify the signature of the credential, as well as the holder's signature, using recorded public keys.
 \end{enumerate}
\end{enumerate}

\begin{figure*}
    \begin{center}
        \includegraphics[width=0.9\textwidth]{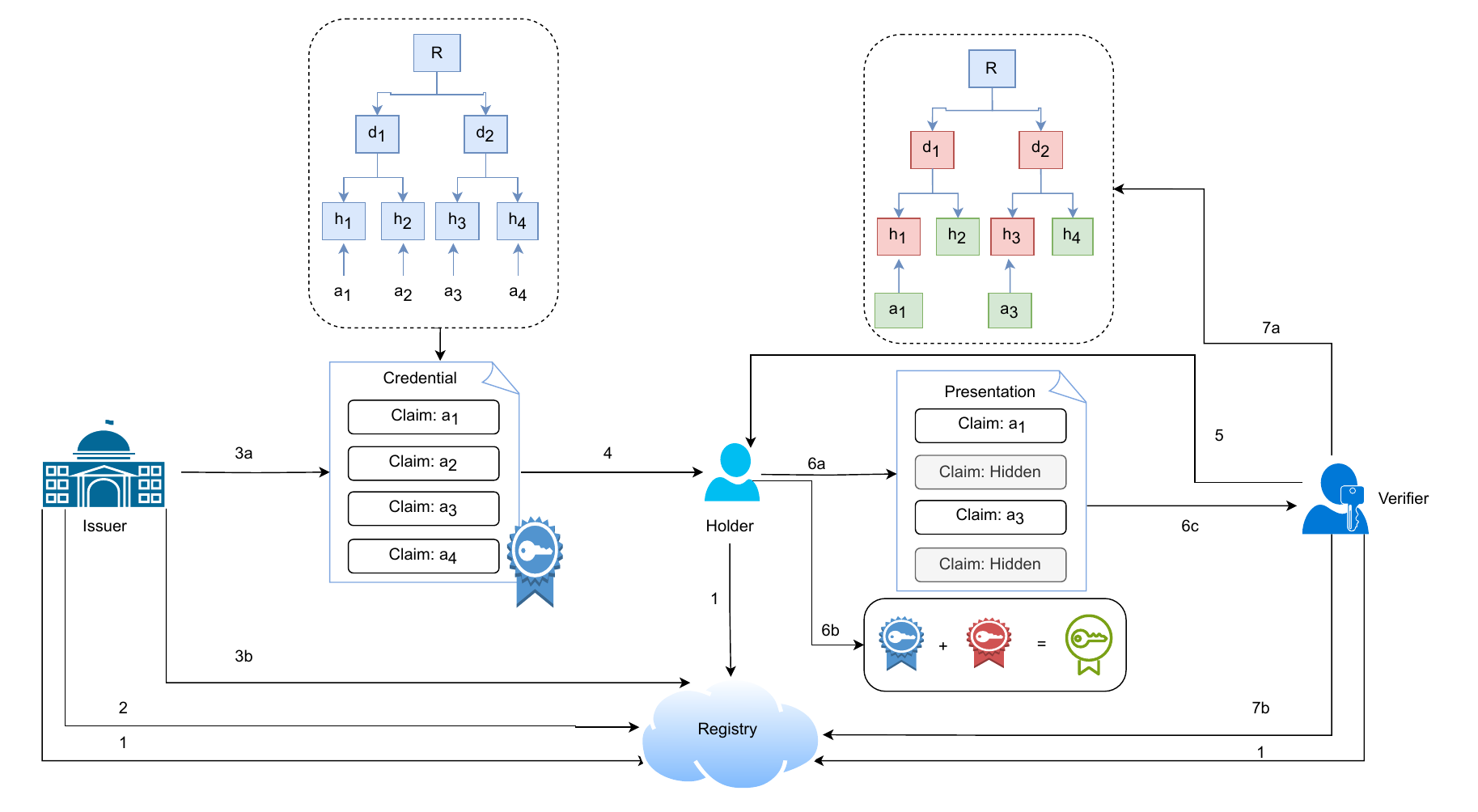}
 \caption{Use case 1: Selective disclosure of claims from a single credential }
 \label{fig:usecase1sol}
\end{center}
\end{figure*}

\subsubsection{Use case 2: Selective disclosure of claims from multiple credentials}
\begin{figure*}
    \begin{center}
        \includegraphics[width=0.9\textwidth]{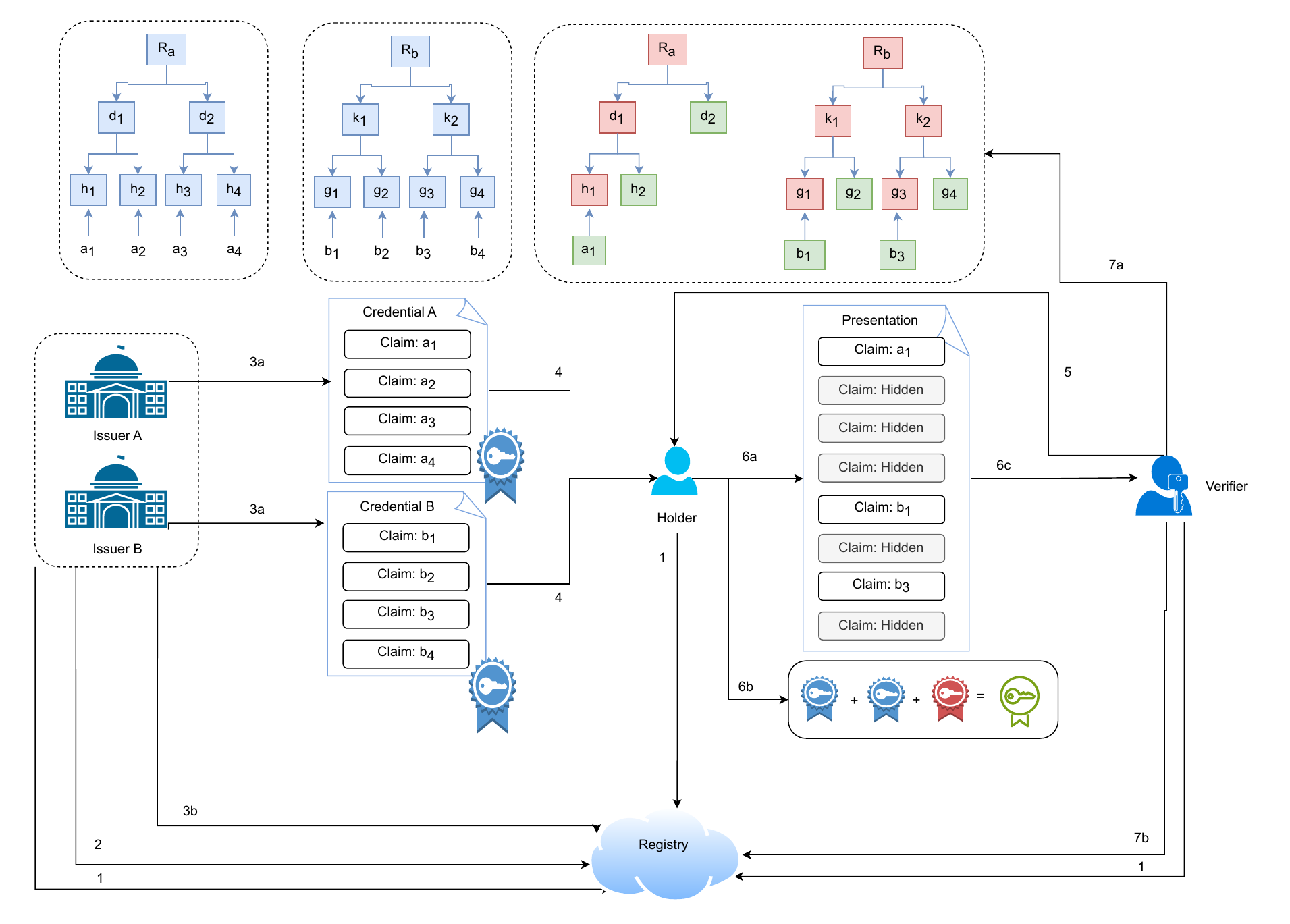}
 \caption{Use case 2: Selective disclosure of claims from multiple credentials}
 \label{fig:usecase2sol}
\end{center}
\end{figure*}

The basis for this use case is BLS signatures that allow the proper merging of multiple credentials. It is shown in Figure \ref{fig:usecase2sol}, and its explanation is given below with the exact steps:

\begin{enumerate}
 \item Entities register their public keys in a public registry;
 \item In addition to the public key, the issuers register the credential format to enable the correct generation of the Merkle tree;
 \item When issuing a credential, the issuers:
 \begin{enumerate}
 \item Create digital credentials and sign the Merkle tree roots using BLS;
 \item Record the issuance proof, i.e. the roots and signatures.
 \end{enumerate}
 \item The holder receives the credentials and stores them in their digital wallet;
 
 \item The verifier requests presentation with specified claims;
\item The holder combines their credentials into a presentation and sends it to a verifier as follows:
 \begin{enumerate}
 \item They reveal claim values that verifier requested along with their salts; for the other values, they prepare the corresponding proofs generated using the Merkle tree. The presentation now consists of all the elements needed to recreate the roots of the Merkle trees;
 \item The signatures of the roots of the trees of individual credentials are aggregated into one presentation signature, together with the holder's signature. This single signature is used to verify the validity of the credentials issuers and holders;
 \item Holder sends a presentation to the verifier.
 \end{enumerate}
 \item When receiving a presentation, the verifier does the following:
 \begin{enumerate}
 \item Recreates the Merkle trees using the registered formats, disclosed values, salts and the resulting hashes, thus obtaining the trees' roots;
 \item Verifies the roots using the public registry (thus validating the credentials' issuance). They verify the aggregated issuers' signature of the credentials and the holder's signature using recorded public keys.
 \end{enumerate}
\end{enumerate}

In this use case, BLS signatures are crucial because they allow the aggregation of signatures; that is, they confirm that the presentation was created based on multiple credentials.

\subsubsection{Use case 3: Selective disclosure of claims/Proving claims without revealing them}

\begin{figure*}
    \begin{center}
        \includegraphics[width=0.9\textwidth]{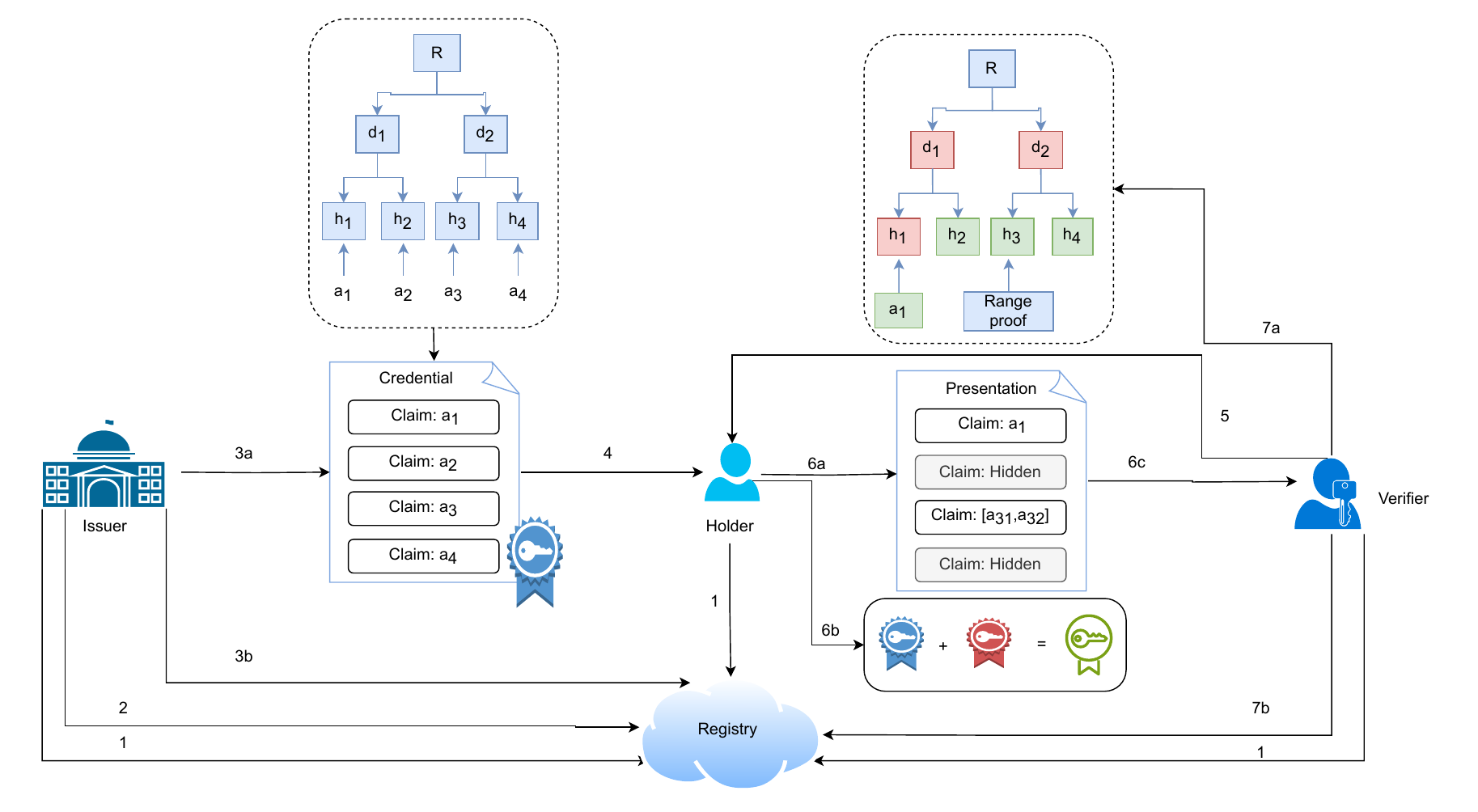}
 \caption{Use case 3: Selective disclosure of claims/Proving claims without revealing them}
 \label{fig:usecase3sol}
\end{center}
\end{figure*}

The third use case extends the previous ones offering range proofs. This feature is crucial in preserving privacy and data minimization. It allows hiding and proving the values instead of revealing and showing them. When implemented, this use case allows users not to reveal their birth date or salary amount but to prove they are in the required range. 
As part of this solution, the key elements are the Pedersen commitments and the ZKP tool Bulletproofs. This tool allows proving that a value belongs to a range, that a certain arithmetic expression is valid, that a value belongs to a set, and so on. The steps of the third use case  (shown in Figure \ref{fig:usecase3sol}) are as follows:
\begin{enumerate}
 \item Entities register their public keys in a public registry;
 \item In addition to the public key, the issuer registers the credential format to enable the correct generation of the Merkle tree;
 \item When issuing a credential, the issuer:
 \begin{enumerate}
 \item Creates a digital credential and signs the Merkle tree root using BLS;
 \item Records the issuance proof, i.e. its root and signature.
 \end{enumerate}
 \item The holder receives the credential and stores it in their digital wallet;
 \item The verifier requests presentation with specified claims;
 \item The holder sends their credential as a presentation to the verifier as follows:
 \begin{enumerate}
 \item They reveal claim values that verifier requested along with their salts; for the other values, they prepare the corresponding proofs generated using the Merkle tree. For values that the holder does not want to reveal but has to prove, they generate a Bulletproof range proof, where they can show that the value belongs to a required range. The presentation now consists of all the elements needed to recreate the root of the Merkle tree;
 \item They sign the root and aggregate their signature with the issuer's signature to create a unique presentation signature;
 \item They send the presentation to the verifier.
 \end{enumerate}
 \item When receiving a presentation, the verifier does the following:
 \begin{enumerate}
 \item Check the range proof validity. Recreates the Merkle tree using the registered format, disclosed values, salts and the resulting hashes, thus obtaining the tree's root;
 \item Verifies the root using the public registry (thus validating the credential issuance). They verify the issuer's signature of the credential, as well as the holder's signature, using recorded public keys.
 \end{enumerate}
\end{enumerate}

In this use case, the critical element is Pedersen's homomorphic commitment and Bulletproofs.

%% file: Chapters/05Results.tex
\section{Time measurements}\label{sec5}

\begin{figure}
    \begin{center}
        \includegraphics[width=\columnwidth]{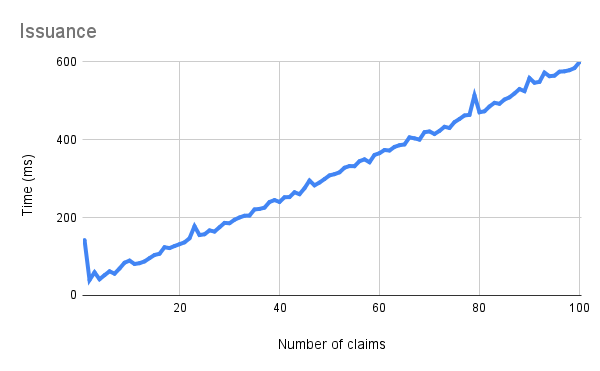} 
        \caption{Credential issuance}
        \label{fig:time1}
    \end{center}
\end{figure}

The proof of concept code of this paper is publicly available on  \href{https://github.com/seilabecirovic/Selective-disclosure-BLS-Merkle-Trees}{Github}. The proof of concept is modularly implemented in JavaScript with separate functions that can be used in plug-and-play manner. To demonstrate how small are the device requirements for this approach , time measurement was conducted on Thinkpad T470 Laptop with Intel® Core™ i5-6300U CPU. The solution's efficiency in issuing many claims, ranging from 1 to 100 with randomized values, is demonstrated and quantified in Figure \ref{fig:time1}. This performance metric clearly explains the solution's capability to manage a large volume of data in short time.

More than 100 claims in a single credential are unmanageable by a holder and should be avoided. It is clear that as the number of claims grows, the time needed to issue them grows, but it is still unnoticeable from a user's perspective.

\begin{figure*}
    \begin{center}
        \includegraphics[width=0.9\textwidth]{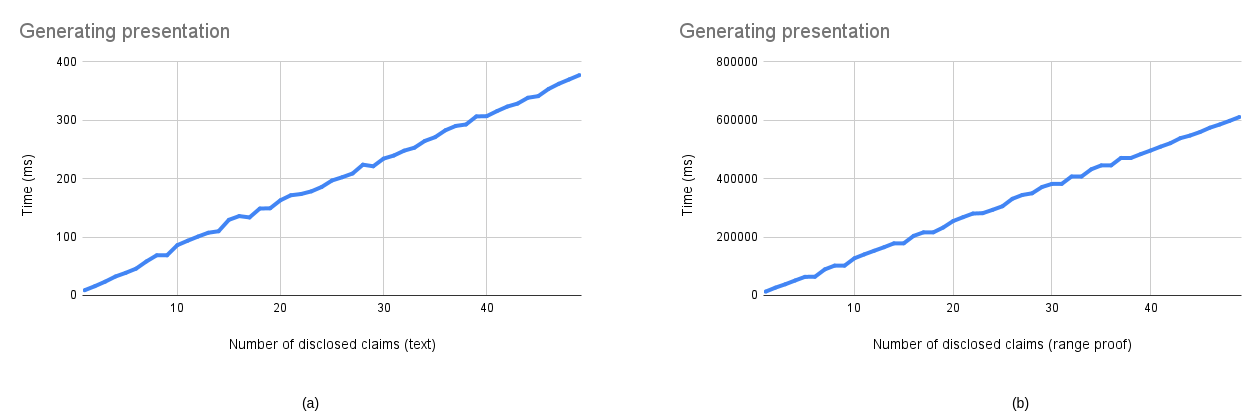} 
        \caption{Generating presentation:(a) Text claims; (b) Range proofs}
        \label{fig:time2}
    \end{center}
\end{figure*}

Proof generation is also measured using a credential containing 100 claims where 1 to 50 were disclosed. It is split into two categories: selective showing of textual claims and selective disclosure of numbers using range proof as shown in Figure \ref{fig:time2}. Selective disclosure using range proof is slower than a simple showing of the attribute, but it can still be expressed in seconds for a small number of range proofs. 

\begin{figure*}
    \begin{center}
        \includegraphics[width=0.9\textwidth]{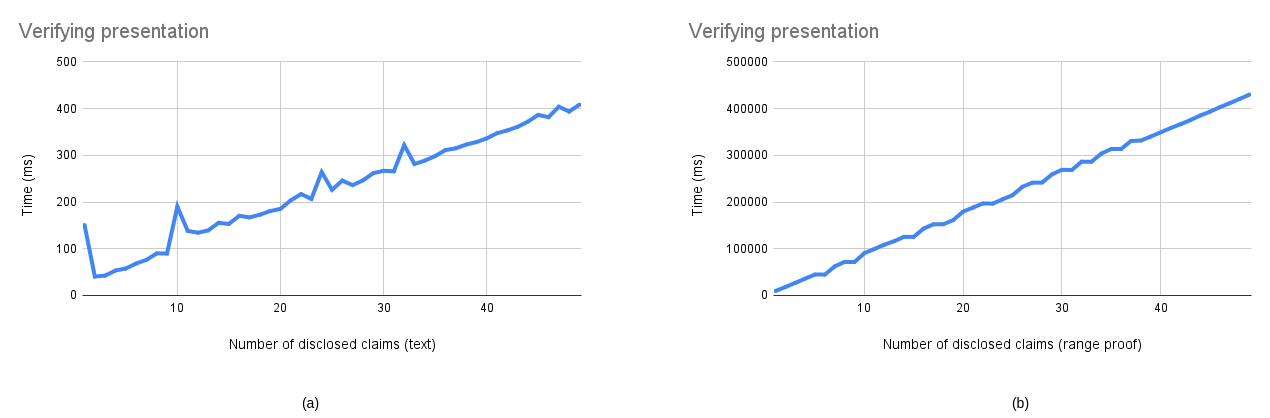} 
        \caption{Verifying presentation: (a) Text claims; (b) Range proofs}
        \label{fig:time3}
    \end{center}
\end{figure*}

Verifying is also split into the same two categories as shown in Figure \ref{fig:time3}, where the time needed to verify range proof is longer than for verification of showed claim. Aggregation of credentials into one and verifying them has only added an element of BLS aggregation, which doesn't affect the time in a significant way.

%% file: Chapters/06Discussion.tex
\section{Discussion} \label{sec6}

The presented solution meets the set requirements and enables different usage scenarios for selective disclosure. It uses:
\begin{itemize}
 \item \textbf{Merkle trees} that allow unlinkability (through hashing and salts), value hiding, and validation of credentials;
 \item \textbf{BLS signatures} that allow verification of issuance, verification of the credentials' holder, combining multiple credentials into one presentation, and the possibility of multiple signatures on one credential;
 \item \textbf{Pedersen Commitments and Bulletproofs} that allow proving values without revealing them in order to promote minimization and overall selective disclosure.
\end{itemize}

\subsection{Comparison to other approaches}

This solution belongs to a new category, and considering that no solution belonging to the same category exists at the time of writing, comparisons from the aspect of work and performance are difficult to make. Nonetheless, in order to present the benefits of this approach, a comparison to existing combined approaches needs to be made (comparing with a single method approach is unnecessary due to additional functionalities achieved through combination):
\begin{itemize}
\item Compared to the combination of hash- and signature-based methods, presented approach enables proving the value of an attribute without revealing it. It also does not require canonicalization algorithms because a single message is signed. Compared to:
\begin{itemize}
\item \cite{39}, presented approach enables combining credentials into a single presentation (req. 2 and 3), proving the value of an attribute without revealing it (req. 12), proving that the identity owner is the subject of the credential (req. 8 and 9);
\item \cite{25}, presented approach enables combining credentials into a single presentation (req. 2 and 3), proving the value of an attribute without revealing it (req. 12), proving that the identity owner is the subject of the credential (req. 8 and 9), uses Merkle tree proofs for faster verification (approach in \cite{25} reconstructs the tree using first layer of tree instead of parent nodes when required).
\end{itemize}
\item In comparison to the hash- and ZKP-based methods \cite{51}, presented approach allows verification of the issuer and identity owner and does not require a trusted setup that is necessary for existing solutions based on zk-SNARKS and proof size is smaller. Compared to \cite{51}, presented approach does not require sending each credential as a separate presentation while using the same salts for attributes (req. 2 and 3).
\item Compared to the signature and ZKP-based methods, it enables explicit credential validation and combining credentials and does not require canonicalization algorithms or a trusted setup. Compared to
\begin{itemize}
\item \cite{31}, presented approach enables combining credentials from different issuers into a single presentation (req. 3),  proving the property of an attribute without revealing it (req. 12);
\item \cite{20}, presented approach enables combining credentials from different issuers or issuers groups into a single presentation (req. 3),  proving the property  of an attribute without revealing it (req. 12);
\item \cite{10}, presented approach enables combining credentials from different issuers or issuers groups into a single presentation (req. 3),  proving the property  of an attribute without revealing it (predicates) (req. 12). Paper \cite{10} is more focused on access control attributes (rules for access control) compared to regular credentials;
\item \cite{43}, presented approach enables combining credentials into a single presentation (req. 2 and 3), proving the property of an attribute without revealing it  (req. 12). The paper \cite{43} focuses on issuer-hiding using ZKP, which contradicts the transparency requirements of digital identity systems in terms of widespread adoption (legal and regulatory).
\end{itemize}
\end{itemize}

\subsection{Performance analysis}
A brief explanation of performance using $\mathcal{O}$ or defined times within the standard,  for individual elements is as follows \cite{becker2008merkle,ietf_bls,tarilab}:
\begin{itemize}
    \item Merkle trees:
    \begin{itemize}
        \item Space - $\mathcal{O}(n)$ 
        \item Searching - $\mathcal{O}(\log{}n)$ 
        \item Traversal - $\mathcal{O}(n)$ 
        \item Insertion - $\mathcal{O}(\log{}n)$ 
        \item Deletion - $\mathcal{O}(\log{}n)$ 
        \item Synchronization - $\mathcal{O}(\log{}n)$ 
        \item Proof -  $\mathcal{O}(\log_2 n)$ 
    \end{itemize}
     \item BLS signatures:
     \begin{itemize}
         \item Private key - 32 bytes
         \item Public key - 48 bytes
         \item Signature - 96 bytes 
         \item Aggregated signature - 96 bytes 
         \item Signing - 370$\mu$ s 
         \item Verifying - 2700$\mu$ s 
     \end{itemize}
     \item Bulletproofs:
     \begin{itemize}
         \item Proof generation - $\mathcal{O}(n)$ 
         \item Proof verification - $\mathcal{O}(n)$ 
         \item Proof size - $3\log{2}n+9$ elements 
     \end{itemize}
\end{itemize}

Due to the sequential nature of the proposed approach, elements such as credential issuance, presentation generation, and presentation verification are complex as the most complex element in them, e.q. Merkle tree generation is the most complex for credential issuance, proof generation and verification from Bulletproofs are the most complex for presentation generation and verification. They are as follows for the three algorithms:
\begin{itemize}
    \item Credential issuance - $\mathcal{O}(n\log{}k)$ - based on generating Merkle tree and Pedersen commitments of attributes that depend on the number of bits $k$ of each attribute value;
    \item Presentation generation - $\mathcal{O}(m*n)$ - based on generating Bulletproofs range proofs for $m$ attributes;
    \item Presentation verification - $\mathcal{O}(m*n)$ - based on verifying range proofs for $m$ attributes.
\end{itemize}
In case of combining credentials, the last two complexities are multiplied by the number of credentials. 

The implemented approach demonstrates not only its usability in practical scenarios and efficient time execution but also its practical application in any format of digital credentials, particularly verifiable credentials. It should be noted that the solution is written in JavaScript and that the time measured depends on the language and the device used. Still, when implemented in the language commonly used for web and mobile applications, the measured time shows that this solution is feasible and viable for practical implementation. 

\subsection{Security and threat analysis}

As this approach uses three primitives in a sequential manner, the proof of security and threats directly depend on the security assumptions and threats of the primitives.

In their basic form, Merkle trees can be attacked in two ways. The first way is if salt is not used. This enables an attack by a rainbow table (tables of already known hashed values). Correlation attacks can also be prevented if different salts are used for different values. 
The root does not reveal the depth of the tree, so Merkle trees are vulnerable to second-preimage attacks (allowing the creation of a Merkle tree with the same root). In order to prevent this attack, it is recommended to enable certificate transparency (adding bits to hashed data and internal nodes) or to limit the tree size \cite{andreeva2009herding,rfc6962}.

The primary advantage of BLS signatures is the possibility of aggregation. In the original BLS scheme, aggregation was vulnerable to rogue public-key attacks. To avoid this, it is recommended to use proof of knowledge of the secret key (KOSK) or unique messages. Boneh, Drijvers, and Neven presented a modified implementation in their work \cite{boneh2018compact}, which is not susceptible to this attack.

Bulletproofs, as presented in the original paper, are susceptible to the Frozen Heart attack. In the case of an insecure protocol, this attack allows falsifying proofs that will still be successfully verified. The original paper uses the Fiat-Shamir transformation to make the proof fully non-interactive. However, this implementation omits a crucial component. To prevent this attack, adding a Pedersen commitment to the Fiat-Shamir transformation hash is sufficient. This prevents proof falsification \cite{trail}.

Another threat to this approach is a man-in-the-middle attack on presentations. A malicious party could intercept and reuse a presentation, potentially spoofing the owner's identity. To mitigate this attack, it is necessary to have nonces, session-based presentation keys, or challenge-response protocols. All of this makes intercepted data unusable in subsequent sessions. Implementations of mentioned elements depends on the system where this approach will be used. 

When designing and implementing a protocol that uses presented approach for selective disclosure, the proposed corrections must be applied to ensure adequate security.

\subsection{Limitations}

The presented selective disclosure approach satisfies the defined requirements but has certain limitations.

One requirement that represents a limitation of the system is unlinkability and collusion resistance. Unlinkability prevents linking presentations of the same user without their permission.The following unlinkability types exist \cite{fett_2022_selective}:
\begin{itemize}
 \item Unlinkability of presentations: The verifier cannot link two presentations of the same credential;
 \item Unlinkability of verifiers:  Two colluding verifiers should not be able to learn that they have received presentations of the same credential;
 \item Issuer and verifier unlinkability (honest verifier): The issuer must not be able to know that the credential was sent to a verifier;
 \item Issuer and verifier unlinkability (compromised verifier): The issuer must not know that the credential was sent to the verifier, even if the verifier tries colluding with the issuer.
\end{itemize}

In all these situations, unlinkability is limited to use cases when the credential does not contain information that directly or indirectly identifies the user, such as a unique ID number or tax number. 
This requirement contradicts the creation of an open and transparent digital identity system.

We will consider the mentioned types of unlinkability from the aspect of the proposed approach:
\begin{itemize}
 \item Unlinkability of presentation and unlinkability of verifiers: By using Pedersen binding values, which are homomorphic, it is possible to prove that a value exists within a credential without revealing it. This does not allow the verifier to know the details of the Merkle tree. For credentials created by the identity owner, they can create a presentation with different salts when creating the tree. In addition, the proposed approach is: When issuing a credential, multiple versions are generated, which differ by salts and, thus, by resultant roots. When an identity owner sends a presentation, a different version is randomly sent each time. In this way, the verifier/s is/are never able to recreate the correct Merkle tree;
 \item Unlinkability of issuers and honest verifiers is not achieved through this approach. However, there is a potential solution to the problem. This solution is of a systemic and regulatory nature. There must be legal frameworks that regulate the honesty and correctness of issuers, as well as properly defined procedures;
 \item Unlinkability of issuers and compromised verifiers is difficult to achieve through the presented approach due to the use of a salt prepared by the issuer. A potential way to prevent collusion is to use the mentioned homomorphism. The identity owner can create appropriate commitments using their salts for attributes whose values can be proven to the issuer. By using ``out-of-the-box'' commitments, the issuer continues with the credential issuing process, and in case of collusion cannot reveal the attributes without knowing salts.
\end{itemize}

Another limitation of the approach is its use in existing or new identity systems. In order to use the approach correctly, it is necessary to:
\begin{itemize}
 \item Fulfill the recommendations of the implementation of individual elements in order to ensure the security of the system;
 \item Fulfill the recommendations of the implementation of individual elements to ensure the performance of the system;
 \item Choose technology and individual elements to achieve the best possible performance.
\end{itemize}

This approach was created in an agnostic manner, meaning it can be used with different credential types: verifiable credentials, attribute-based credentials, and anonymous credentials. It can also be used with different identity models, from centralized and federated to decentralized and self-sovereign identity models. In the case of the centralized and federated identity model, the recording of the credential issuance in the public registry refers to the recording of the issuance in the corresponding centralized database. The step in which the verifier validates and verifies the presentation is also done through communication with the issuer (without reliance on a public data registry).

The performance and security of this approach depend on the implementation of the entire system, which means that this approach, implemented within the system, increases in complexity. Depending on the identity model used, key management, recording of credentials issuance, and verification processes can change. For example, in centralized or federated models, PKI (public key infrastructure) can be used, while in decentralized models, DIDs (decentralized identifiers) and DID documents can be used. 

Integrating this approach with existing identity models is necessary for its widespread adoption. It is necessary to define how this approach can be layered onto existing identity infrastructures without disruption. Key elements that need to be considered are scalability, trust management and user experience. Optimizing each element of the presented approach will allow for usage in high-volume environments. Using standardized cryptographic primitives and implementation will allow regulatory compliance in centralized and federated models. Usability and user interaction for this approach are also key to the adaptability of the approach and one of the challenges in practical implementation. User experience managing the credentials and creating presentations must be considered when implementing the approach.  

This approach requires standardizing certain elements. It is necessary to define the format of required claims, how the proofs will be added to verifiable presentations, and how the revocation of issued credentials can be done. Each of these elements can affect the widespread adoption of this approach. Even though the presented approach fulfills the requirements, standardization efforts may face different challenges. These challenges include interoperability across implementation, which depends on the exact implementation of each element used, and various regulations, which are different depending on the jurisdiction and evolving landscape, where techniques and digital identity standards are rapidly evolving. Each implementation of this approach should be interoperable, and this can be achieved by defining precise specifications and reference and test implementations. This approach is modular and adaptable to address regulatory differences in case of a requirement for specific cryptographic primitives. To achieve proper standardization, it is necessary to update this approach to be compatible with emerging standards such as W3C VCs and DIDs.

\subsection{Future work and potential improvements}

Using this approach, it is possible to change some primitives to achieve adequate security. Moreover, one of the future challenges is to verify and compare the behaviour of similar algorithms in the approach or to improve the algorithms to those that can be executed in the post-quantum world. For example, it is possible to replace Bulletproofs with some zk-STARK algorithm. Both variants do not require a trusted setup. Although zk-STARK is not widely used in this field \cite{oudesystematic}, current performance and the existence of post-quantum tools would allow new elements. In addition, Pedersen binding values can be replaced with the Poseidon hash \cite{grassi2021poseidon}, which has been proven adequate for zk-STARK tools. In contrast, BLS signatures can be replaced with other aggregateable signatures, such as PQScale \cite{hsiang2023pqscale} - FALCON \cite{alagic2022status}. In order to reduce proofs in presentation, it is also necessary to consider aggregating Merkle tree proofs.

%% file: Chapters/07Conclusion.tex
\section{Conclusion}\label{sec7}
In this paper, we presented a solution for the selective disclosure of claims in digital credentials. By combining Merkle hash trees with BLS signatures and the ZKP method Bulletproof, we fulfilled the requirements of selective disclosure. This solution allows for:
\begin{itemize}
\item Generation of credential with multiple issuers, which has only one aggregated signature;
\item Generation of credential that both issuer and holder signs, which has only one aggregated signature;
\item Selectively disclosing claims from a single credential, while ensuring unlinkability and maintaining the verifiable pairing between the holder and the credential;
\item Selectively disclosing claims from multiple credentials, combining them into one presentation, while ensuring unlinkability and maintaining the verifiable pairing between the holder and the credential;
\item Generating small range proofs for values that should not be disclosed and verifying them. 
\end{itemize}
The solution presented belongs to a category of combined approaches for selective disclosure: ZKP, hash- and signature-based. The solution is usable and practical in real-life scenarios. 
Future work will include creating verifiable credentials using this approach for a self-sovereign identity system. It should also consider other ZKP tools that can be used and the possibility of a post-quantum solution that uses this approach but different methods. 
New approaches to selective disclosure are needed as the rapid development of digital identity systems demands enhanced privacy and security mechanisms to keep pace with evolving technological and user needs.